# Relaxed enthalpy and volume during physical aging: Their interrelation

Vassiliki Katsika-Tsigourakou[*], Georgios E. Zardas

*Section of Solid State Physics, Department of Physics, National and Kapodistrian University of Athens, Panepistimiopolis, 157 84 Zografos, Greece*

**Abstract**

Several papers have recently presented results of measurements of physical aging by studying the behavior of glassy materials quenched from temperatures above their glass transition temperature $T_g$. The evolution of the aging process is usually followed by plotting the relaxed enthalpy versus the accompanying decrease in volume. Here, we focus on the slope of such plots, which are found to be similar to the inverse value of the isothermal compressibility close to $T_g$. An explanation of this empirical result is attempted in the frame of a model that interconnects the defect Gibbs energy with properties of the bulk material.

-----------------------------------

[*] vkatsik@phys.uoa.gr

# 1. Introduction

Time dependent phenomena referred to as physical aging are normally associated with the behavior of amorphous polymers or other glass-formers quenched from temperatures above their glass transition temperature, $T_g$ (e.g., see Refs [1,2]). When stopping the cooling procedure at a temperature below $T_g$ and keeping this temperature constant, one measures a continued decrease in volume, $v$, which is just an example of a physical aging phenomenon. This is accompanied by an amount of heat leaving the sample after the cooling procedure, which is in fact the relaxed enthalpy $h$. The relaxed enthalpy versus volume graphs have been the object of several recent experimental studies. As a first recent example, we refer to the detailed study of Slobodian et al [2], who measured a number of polymers (PMMA, PMMA/PEO blends, PS, PC, PVC, PET) and amorphous selenium. Their aging temperature was -15°C and the slope $dh/dv$ was determined to be around 2 GPa for the polymers, while for Se a higher value of 4.9 GPa was found. As a second example, we mention the most recent study by Hadac et al [1] of the effect of cooling rate on the enthalpy and volume relaxation of polysterene (PS). This is the first study [1] that includes results from enthalpy and volume relaxation measurements performed on PS calorimetric and dilatometric samples with same thermal history, cooled at a constant rate from equilibrium at above $T_g$ to the aging temperature $T_a$. Their $h(v)$ graphs led to a slope $dh/dv$ around 1.8 GPa.

A point of chief importance that emerged from the aforementioned experimental studies is the following: Comparing the slopes $dh/dv$ with directly measured compressibility ($\kappa$) data, a close similarity between $dh/dv$ and $1/\kappa$ was found [1,2].

The interpretation of this important empirical fact has recently attracted a strong interest (see Ref. [1] and references therein). It is the object of this short paper to raise the possibility that this could be understood in the frame of the so called $cB\Omega$-model (see below) that has long been suggested for the interconnection of point defect parameters in crystalline solids with bulk elastic and expansivity data. In the next Section we briefly summarize the aspects of this model and in Section 3 we indicate a possible interpretation of the empirical fact under discussion.

**2. The model that interconnects point defect parameters with bulk properties**

The model (termed $cB\Omega$-model) suggests that the defect Gibbs energy $g^i$ is interrelated with the (isothermal) bulk modulus $B(=1/\kappa)$ and the mean volume per atom $\Omega$ through the relation [6-10];

$$g^i = c^i B \Omega \tag{1}$$

where $c^i$ is a dimensionless constant, almost independent of temperature and pressure, and the superscript $i$ corresponds to the defect process (e.g., formation, migration, activation etc) under consideration. The defect volume $\upsilon^i$, is found by inserting Eq. (1) into the relation $\upsilon^i = (dg^i/dP)_T$, which leads to [7,8]:

$$\upsilon^i = c^i \Omega \left[ \left( \frac{dB}{dP} \right)_T - 1 \right] \tag{2}$$

where $P$ denotes the pressure. A combination of Eqs (1) and (2) gives:

$$\frac{g^i}{\upsilon^i} = \frac{B}{\left( dB/dP \right)_T - 1} \tag{3}$$

We shall return to this relation later on.

The defect entropy $s^i$ is found by inserting Eq. (1) into $s^i = -(dg^i/dT)_P$, which leads to: $s^i = -c^i[d(B\Omega)/dT]_P$. Combining this relation with the equation $h^i = g^i + Ts^i$, where $h^i$ denotes the defect enthalpy, and using also Eq. (1), we finally get:

$$h^i = c^i \Omega \left[ B - T\beta B - T\left(\frac{dB}{dT}\right)_P \right] \qquad (4)$$

where $\beta$ is the thermal volume expansion coefficient. Equation (4), in conjuction with Eq. (2), gives:

$$\frac{h^i}{\upsilon^i} = \frac{B - T\beta B - T\left(dB/dT\right)_P}{\left(dB/dP\right)_T - 1} \qquad (5)$$

The compatibility of the $cB\Omega$-model with experimental data has been already checked for the thermodynamic parameters related to the defect formation and migration processes in a variety of crystalline solids [6-10]. In addition, the following three important experimental facts have been explained. First, when measuring the ionic conductivity ($\sigma$) in alkali and silver halides at temperatures ($T$) close to the melting temperature, the plot $\ell n(\sigma T)$ versus $1/T$ exhibits an upward curvature. This was found to be reproduced [11,12] by means of Eq. (1). Second, when applying uniaxial stress on ionic crystals, electric signals are emitted due to the formation and migration of defects that have parameters consistent with the $cB\Omega$-model [13]. The emission of these signals play an important role in the understanding of the low frequency electric signals that are detected before major earthquakes [14-18]. Third, the ionic conductivity (and diffusivity) in mixed alkali halides [19,20] exhibit a non linear variation with composition that agrees with the predictions of the $cB\Omega$-model.

The same holds for their dielectric constant [21], the volume dependence of which can be expressed through the bulk modulus $B$ [22].

In view of the above confirmations of the $cB\Omega$-model we now proceed, in the next Section, to discuss its applicability to the case of the relaxed enthalpy vs volume plots.

**3. The enthalpy and the volume relaxation in glass-formers**

The measurements carried out after cooling the samples at constant rate to the aging temperature, show [1] that the plots $h$ vs $\ell ogt$ and $\upsilon$ vs $\ell ogt$ are linear over several decades of the (elapsed) time $t$. This, which implies the exponential dependences: $dh/dt = \exp(Ch)$ and $d\upsilon/dt = \exp(C'\upsilon)$, establishes [1] a natural link to the concept of thermally activated processes, thus allowing the application of defect activation models to the present case.

Let us start from Eq. (3), which whenever $Ts^i << h^i$ and therefrom $g^i \approx h^i$, can be approximately written as:

$$\frac{h^i}{\upsilon^i} \approx \frac{B}{\left(dB/dP\right)_T - 1} \tag{6}$$

The denominator of the right hand side is practically temperature independent since in crystalline solids it is related to Born's exponents [23]. Furthermore, following very recent aspects [24], we apply a rule-of-thumb stating that: $dB/dP = 5$. Hence Eq. (6) now reads:

$$\upsilon^i = 4(1/B)h^i \tag{7}$$

This, which does not have any adjustable parameter, strikingly coincides with the empirical relation of Hadac et al. [1] (see their Equation (12)) that was found to describe, at least qualitatively, the experimental data (if we assume of course that the defect parameters $\upsilon^i$ and $h^i$ in Eq. (7) correspond to the quantities $\upsilon$ and $h$ measured in Ref. [1]).

A more accurate calculation requires the use of Eq. (5), but unfortunately some of the elastic and expansivity data involved in its right hand side are not yet available to us.

## 4. Conclusion

Using the $cB\Omega$-model, that interconnects point defect parameters with bulk elastic and expansivity data, we can successfully reproduce an empirical relation, i.e., Eq. (7), that was found [1,2] to describe the experimental data.